\documentclass{article}
\usepackage[utf8]{inputenc}

\usepackage{graphics}
\usepackage{graphicx}
\usepackage{mathtools}

\usepackage{geometry}

\usepackage{makecell}
\usepackage{multirow} 
\usepackage[table]{xcolor} 

\usepackage{caption}
\usepackage{graphicx}%

\usepackage{siunitx}
\DeclareSIUnit\bar{bar}
 
\usepackage{parskip}
\usepackage{amsmath,amssymb,amsfonts}%
\usepackage{amsthm}%
\usepackage{mathrsfs}%
\usepackage[title]{appendix}%
\usepackage{textcomp}%
\usepackage{manyfoot}%
\usepackage{booktabs}%
\usepackage{algorithm}%
\usepackage{algorithmicx}%
\usepackage{algpseudocode}%
\usepackage{listings}%
\usepackage[version=4]{mhchem}

\usepackage[allcolors=false, hidelinks]{hyperref}

\newcommand{\adj}[1]{{\color{black} #1}}
\newcommand{\adjJ}[1]{{\color{black} #1}}

\newcommand{\esc}[1]{{\color{black} #1}}

\begin{document}

\title{Controlled nucleation in methylamine-treated perovskite films by artificial seeding and phase-field simulations}

\maketitle

Emilia R. Schütz$^{a\dagger}$, Martin Majewski$^{b\dagger}$, Olivier J.J. Ronsin$^b$, Jens Harting$^{b,c}$*, Lukas Schmidt-Mende$^a$*\\

a) Department of Physics, University of Konstanz, Universitätsstraße 10, Konstanz, 78464, Germany\\

b) Helmholtz Institute Erlangen-Nürnberg for Renewable Energy (HIERN), Forschungszentrum Jülich GmbH\\

c) Department of Chemical and Biological Engineering and Department of Physics, Friedrich-Alexander-Universität Erlangen-Nürnberg, Fürther Straße 248, Nürnberg\\

$\dagger$ These authors contributed equally to this work.

Email Address: j.harting@fz-juelich.de, lukas.schmidt-mende@uni-konstanz.de\\

\abstract{
Large perovskite crystals with reduced defect density enable superior charge transport and stability. Therefore, controlling their nucleation and growth is key to advancing high-performance optoelectronic devices based on perovskite semiconductors. Millimeter-scale perovskite crystals can be synthesized as a continuous film through methylamine treatment, with nucleation sites directed by pre-patterned seeds. Nonetheless, certain configurations may lead to unwanted parasitic nucleation. To predict and mitigate this effect, we employ phase-field simulations alongside an analytical model. Their predictive capability is demonstrated across three distinct material–substrate systems, enabling precise control over nucleation and subsequent crystal growth. Notably, the only material-specific input required is the nucleation density (i.e., the number of crystals nucleated per unit area on an unpatterned substrate). This generality makes the models broadly applicable to diverse material systems for achieving controlled two-dimensional crystallization for improved optoelectronic device performance.
}
keywords: methylamine treatment, seeded growth, PF simulation, analytical model

\section{Introduction}\label{sec1}

Perovskite materials \adjJ{attracted} significant attention in recent years due to their remarkable optoelectronic properties  \cite{perovskitesareawesome_Yang2024}. The formation of large perovskite crystals, ideally to single-crystalline devices, is considered highly desirable, as it reduces defects at grain boundaries  \cite{GrainBoundariesReview_zhang_2024} and enhances the stability of the material  \cite{singleCrystal_review_2024}. These characteristics are crucial for improving the performance and longevity of perovskite-based devices.\\
An effective method to achieve large grain sizes in perovskite films is through methylamine treatment. Through the reaction between methylamine gas and the perovskite \adjJ{material}, an existing film can be dissolved and recrystallized with altered properties that are determined only by this \adjJ{single} treatment step \cite{healing_origin_Zhou2015}. 
\adj{As the original film quality is not decisive for the final film properties, this treatment potentially eliminates the need for meticulous optimization of earlier steps. In addition, it enables straightforward upscaling by facilitating large-scale fabrication without demanding high precision in the initial film deposition.}
Crucially, the method can be used to \adjJ{slow} down a nucleation and growth process, producing exceptionally large grains \adjJ{that} are beneficial for applications in solar cells  \cite{healing_devices_Fan2020, Jacobs2016SI} and photodetectors  \cite{Schtz2022ReducedDD}.\\
The ability to pattern substrates with artificial nuclei offers further control over the crystallization process, allowing for the pre-determination of grain size, position, and orientation  \cite{Au_dots_geske2017, guenzler_FIRA_2021}. This level of control could be particularly advantageous for applications requiring high crystallinity and precise structural organization, such as in LEDs, laser devices, and lateral perovskite devices \adjJ{such as} back-contacted solar cells. Moreover, patterned growth \adjJ{enables us to create} interesting model systems to deepen our understanding of the crystallization process itself.\\
Controlled two-dimensional growth has been explored in various material systems, and its broader implications extend beyond perovskites  \cite{kapadia2014deterministic, miseikis2017deterministic, epitaxy_wang2016chemically, seeded_epitaxy_xu2021seeded, crystallization_transistors_yu2017}. The ability to grow highly crystalline elements directly on circuit structures with controlled properties opens up possibilities for numerous technological applications. However, despite its potential, there has been limited research into understanding and simulating these types of growth processes. Much of the current knowledge is based on trial and error, underscoring the need for a more systematic investigation into the limitations and capabilities of patterned growth techniques.\\
To go beyond trial and error, a theoretical description of the underlying processes is required.
\adjJ{Phase-field (PF) simulations offer an attractive approach to this problem, since they can} numerically describe the phase change from amorphous to solid  \cite{granasy_modelling_2004,wu_phase_2021,warren_phase_2009}. PF models are coarse-grained models with diffuse interfaces, based on the minimization of a free energy. 
The phenomenon of nucleation can be included in PF models by applying fluctuations on the evolution equation \cite{shen_effect_2007}.
By including this effect, heterogeneous nucleation and crystal growth in confined spaces have been investigated  \cite{granasy_phase-field_2019}. The solution processing of perovskite systems has been investigated by studying the interplay of solvent evaporation and crystal growth \cite{starger_quasi-2d_2023} and furthermore including nucleation \cite{majewski_simulation_2024,qiu_over_2025}. 
Nevertheless, to the best of our knowledge, the presence of seed crystals in the initial state and allowing additionally for spontaneous nucleation has not been investigated in the literature so far.\\ 

\adjJ{In this work, two of these models, a phase field simulation and an analytical model, are used to optimize the  experimental patterning of a methylamine-treated perovskite thin film.}
\adj{ The main problem with patterning is the appearance of '\adjJ{parasitic}' crystals at spots which are not intended by the patterning. We recover this behavior with PF simulations \cite{ronsin_phase-field_2022}, where we place crystals in the initial state in an hexagonal lattice and allow for additional nucleation. Based on the simulations, we propose an analytical model describing this behavior. Our models show that the main criteria for the appearance of parasitic crystals is \adjJ{an dimensionless parameter, calculated from the} distance between the seeds and the nucleation density. \adjJ{Consequently}, we predict this distance depending only on the nucleation density of the system. \adjJ{Hence, we are able to evaluate the feasibility of any arrangement of seeds to fully control the nucleation and growth process and avoid parasitic crystallization.}

The paper is structured as follows: First, we introduce the methylamine treatment and the patterning processes. Then, we describe the theoretical results including the phase-field simulations, the derivation of the analytical model, and the validation of the predictions for three different experimental systems. Finally, the generality of the models is discussed before we conclude with a summary of our finding.}
\\

\section{Methylamine treatment process}\label{secMethylamine}
\begin{figure}[h!]
\centering
\includegraphics[width = 0.75\textwidth]{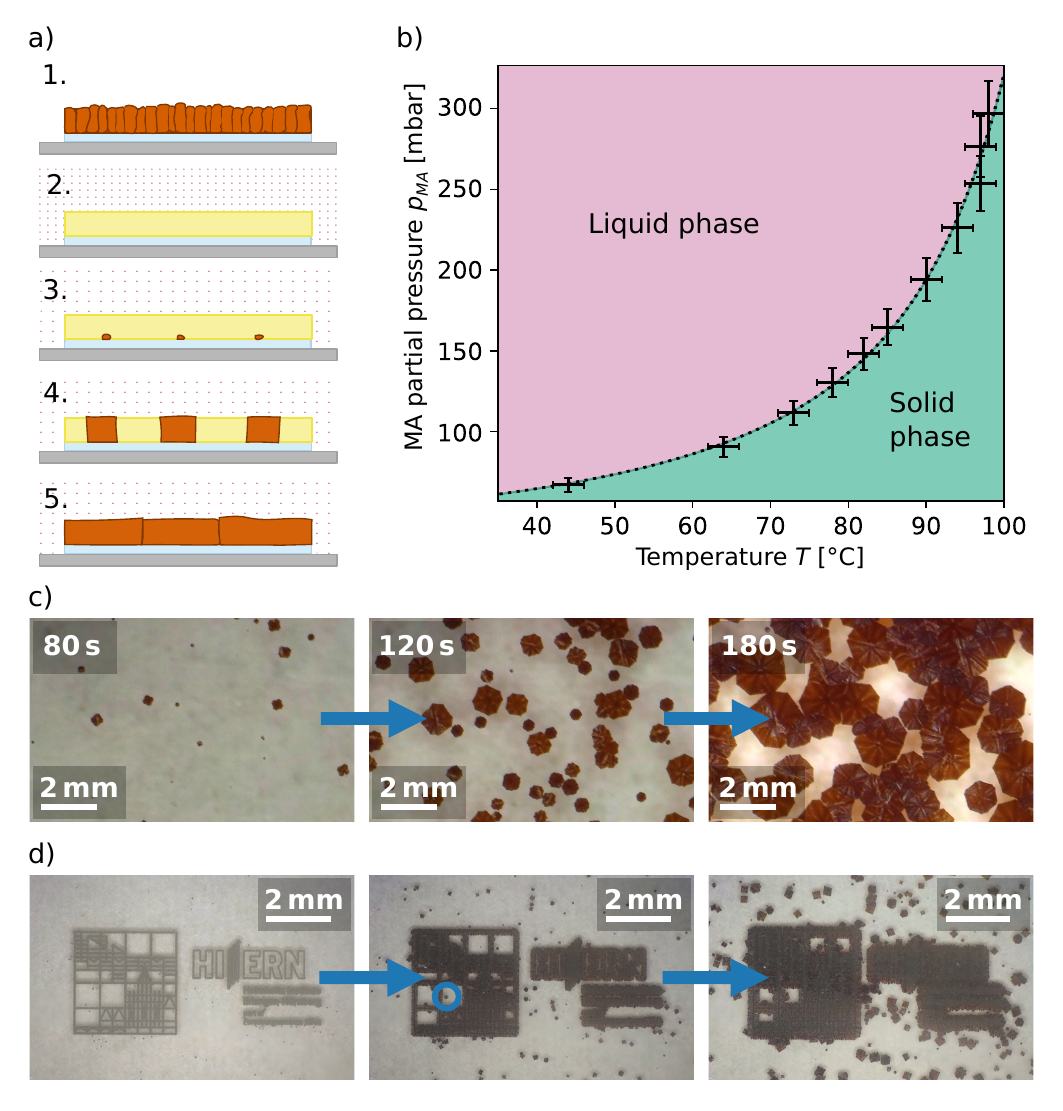}
\caption{
a) Schematic of the methylamine treatment process. b) Phase diagram of the methylamine process: For sufficiently high methylamine \adjJ{(MA)} partial pressures $p_{MA}$ at sufficiently low temperatures $T$, the system enters a liquid intermediate state. Through a reduction in $p_{MA}$, the system is brought just across the phase boundary, where slow nucleation and recrystallization ensues. The quantitative values presented here were derived for triple cation perovskite on an ITO substrate. c) Images of a sample during recrystallization. d) Images of the successful artificial nucleation of the perovskite. Note that the underlying Au structure is not visible, only the resulting perovskite crystals. Where the nucleating structures are too far apart, additional  'parasitic' nuclei grow (an example is circled in blue).
}
\label{fig:TreatmentProcess}
\end{figure}

The methylamine post-treatment we apply in this work is a versatile instrument in modifying perovksite thin films \cite{healing_origin_Zhou2015} and achieving up to millimeter-sized 
grains  \cite{healing_devices_Fan2020, Schtz2022ReducedDD}. It allows us to achieve controlled, very slow nucleation and crystallization, yielding a compelling model system to observe growth mechanisms.
\\
The full treatment process is shown in \autoref{fig:TreatmentProcess} a). 
It is based on the reaction between methylamine gas and the perovskite lattice: The amine replaces the Pb-I bonds due to its proclivity to coordinate Pb$^{2+}$ \cite{Feng2020}. Consequently, the perovskite lattice collapses into a liquid intermediate state, with the methylamine acting essentially as a gas-phase solvent \cite{Bogachuk2020}:
\begin{equation}
    \ce{ABX_{{3(s)}} + $x$ CH3NH2_{(g)}} \rightleftharpoons \ce{ABX_{3} \cdot $x$ CH3NH2_{(l)}}
\end{equation}
Once the methylamine leaves the film, inducing a state of supersaturation, the perovskite phase recrystallizes. In \autoref{fig:TreatmentProcess} b), the phase diagram of this system is shown. Experimentally, we achieve liquefaction or recrystallization by changing the  system temperature $T$ and 
the methylamine partial pressure $p_{MA}$. 
\autoref{fig:TreatmentProcess} c) shows the nucleation and recrystallization process, as observed in-situ. Grains nucleate at random points, then 
grow outward until they cover the entire surface. For a more detailed explanation of the treatment process, see  \cite{Schtz2022ReducedDD}. 
\\
During this process, the exact grain location, i.e., the points of nucleation, remain \adj{functionally random, as they are likely to be determined by impurities or protrusions on the substrate}. To gain complete control over the final film, we thus introduce artificial seeds onto the substrate.

\subsection{Gold patterning for nucleation control}
Generally, nucleation most likely occurs heterogeneously at the substrate/liquid interface, and the likelihood of nucleation is highly sensitive to the substrate material.
Gold has been found to be an effective seed material for various substances  \cite{yan2016selective, Li_Au_seeding2018}, including perovskites  \cite{Au_dots_geske2017}. By using lithographically patterned gold dots as nucleation centers, we can thus create preferred points for nucleation on our substrate.
By carefully selecting the recrystallization parameters, \adjJ{the temperature} $T$ and \adjJ{the methylamine partial pressure} $p_{MA}$, we maintain a state of supersaturation that prevents nucleation at the substrate interface but allows \adjJ{for} it at the gold sites. This enables us to effectively control the precise arrangement of grains in the film.
For a more detailed explanation of the nucleation control mechanism, see the Supplementary Information, section S1.
In \autoref{fig:TreatmentProcess} d), the controlled nucleation and growth on the gold patterned substrate is shown.
\\
While this scheme delivers the controlled growth of large-grained perovskite structures, there are limits to the seed pitch
\adjJ{at which this process works well}: As there is a remaining nucleation probability at the substrate interface, \adjJ{parasitic} nucleation is observed between the artificial seeds above a certain seed distance $D_{max}$ (see the features circled in blue in \autoref{fig:TreatmentProcess} d))
. The extend to which unwanted, parasitic, grains appear is of great importance for future applications: $D_{max}$ effectively limits the size of any circuit or device into which the grown grains are to be integrated. This question does not only pertain to this specific material system, but to all 2D nucleation systems, in some of which parasitic grain formation is also clearly visible \cite{kapadia2014deterministic}.
In the following, we approach this problem from a theoretical perspective by using phase-field simulations and constructing an analytical model.
\section{Results}\label{sec2}

\subsection{Phase Field predictions for the maximal seeding distance}\label{sec:PhaseFieldSimulationsResults}

\begin{figure}[h!]
\centering
\includegraphics[width = 0.75\textwidth]{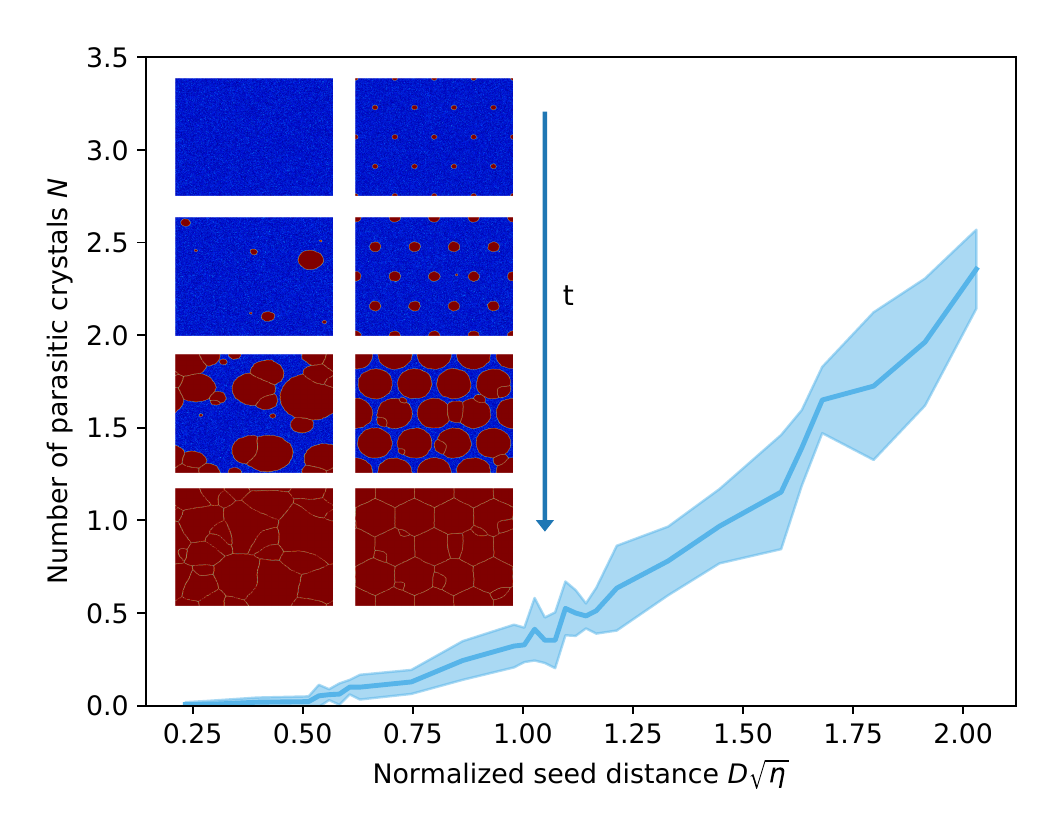}
\caption{The average number of \adjJ{parasitic} crystals in dependence on the pitch D. The shaded area depicts the standard deviation over 10 simulation runs. \adjJ{Note that the only difference between the simulations, for the same pitch, is the random fluctuations.} Insets: Snapshots of the time evolution of a substrate without (left) or with (right) initial seed crystals in the simulation. The amorphous phase is depicted in blue, and the crystalline phase is shown in red.
}
\label{fig:TimetraceSim}
\end{figure}

\adjJ{In this chapter are phase-field (PF) simulations are presented. } 
The equations governing the simulation \adjJ{have been published \cite{ronsin_phase-field_2022}} and are described in the Supplementary Information, section S2. A two-dimensional top view of the system with periodic boundary conditions in both directions. \adjJ{The initial amorphous material transforms to the crystalline phase with the system's evolution described using an order parameter.} To allow for handling multiple crystals, each crystal is uniquely labeled with a marker field. Nucleation is triggered by thermal fluctuations of the order parameter. \\
To mimic the gold patterning of the substrate, the initial state of the simulation features round crystals positioned in a hexagonal lattice. Like in the experiments, we vary the distance between the seed crystals. 
Ten different simulations are performed for each pitch, and the number of \adjJ{parasitic} grains is recorded.\\
The resulting number of parasitic crystals depends on the nucleation rate $\kappa$ and the growth rate of the nucleated crystals $v_g$. These quantities depend on the combination of material and substrate. However, they are often unknown and difficult to measure experimentally. To connect the simulation results to experimental systems, we scale the results to the final crystal density $\eta$ on an unpatterned substrate:
\begin{equation}
    \eta = \frac{\text{Number of crystals}}{\text{Area}}.
\end{equation}
$\eta$ depends on \adjJ{ crystal surface tension $\kappa$ and crystal growth speed $v_g$}, and is (relatively) easily accessible experimentally. The final results depend only on this single material property and the pitch $D$, \adjJ{ which allows for the comparison between simulation and experiments.}\\
The number of \adjJ{parasitic} crystals $N$, averaged over 10 simulation runs and scaled by the number of initial seeds, is shown in \autoref{fig:TimetraceSim}. 
As described above, the pitch is scaled with the square root of the nucleation density $\eta$. 
We observe that there is a range of small pitches, up to $D\sqrt{\eta}$ = 0.5, where (nearly) no parasitic nucleation occurs. \adj{Interestingly, exactly no parasitic nucleation occurs only for smaller values up to $D\sqrt{\eta}$ = 0.25.}
Above $D\sqrt{\eta}$ = 0.5, the number of parasitic grains increases. The results are robust against changes in the sizes of the seed crystals and lattice arrangement (see Supplementary Information, sections S2.5 and S2.6). \adj{To gain a more fundamental insight into the number of parasitic grains, we develop an analytical model of the process in the following section.} 

\subsection{Analytical model}\label{sec:AnalyticalModelResults}

\begin{figure}[h!]
\centering
\includegraphics[width = 0.75\textwidth]{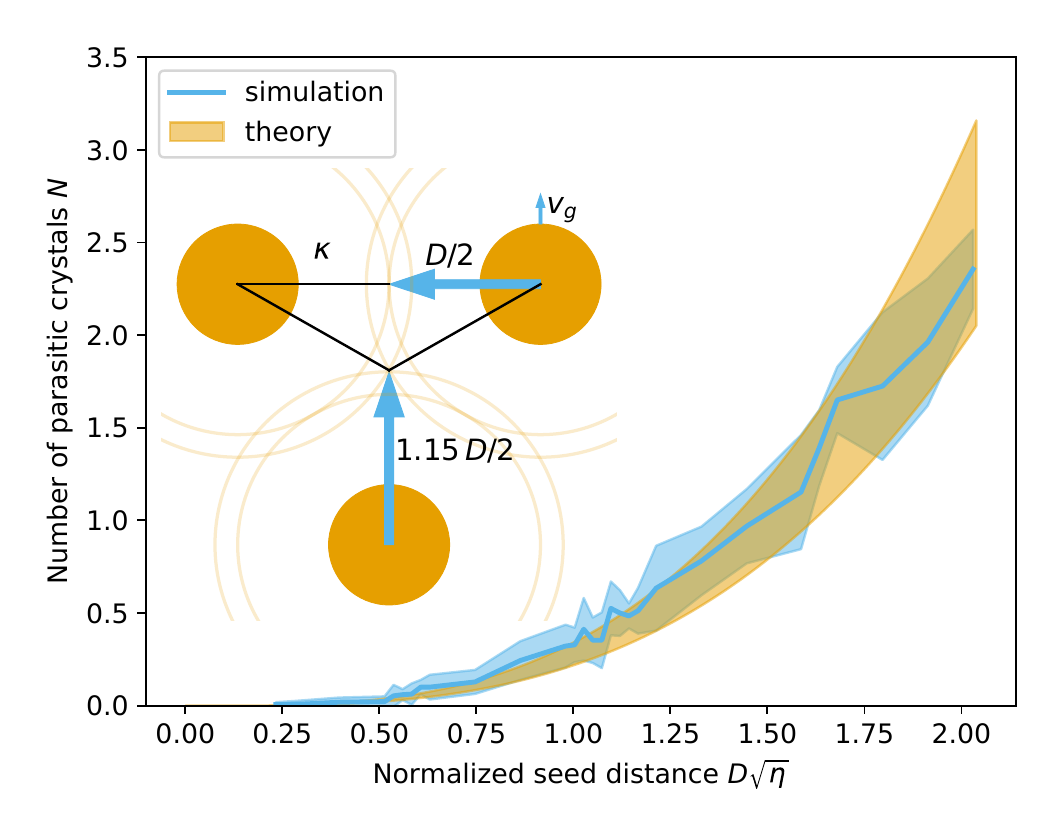}
\caption{The number of \adjJ{parasitic} crystals obtained from the simulation and calculated analytically (\autoref{equ:AnaModel}). Inset: sketch visualizing the parameters of the analytical model. \adjJ{The three crystals (orange) represent an unit cell of the hexagonal lattice. The crystals grow with a growth speed of $v_g$, while additional crystals nucleate in the amorphous phase with a rate of $\kappa$. The arrows and shaded circles visualize the lower and upper integration limits of the model (see main text).}}
\label{fig:AnModel}
\end{figure}

In order to derive the model, we consider a single triangle of \adjJ{crystals. This represents the basic unit of the hexagonal lattice used in the experiments.} The appearing crystals have an initial size of $r_0 = 0$ and growth with a speed of $v_g$. An amorphous precursor phase is present between the crystals. In the amorphous phase, the nucleation rate is $\kappa$, which is assumed to be constant in time and space. \adj{The number of \adj{parasitic} grains $N$} can be calculated by
\begin{equation}
   N = \int_{0}^\infty {\kappa A(t)},
\end{equation}
where $A(t)$ is the surface which is not crystallized at time $t$. \adj{When crystals are placed initially, $A(t)$ has to account for the growth of these crystals. }
For simplicity, we neglect the effect of parasitic crystals on the available volume for crystallization. \adjJ{This will lead to an overestimation of crystals for large seed distances.} We expect to have a nucleation induction time $t_0$, which is the time between the start of crystal growth and the first nucleation event. Since the crystal growth of the seeds will result in full coverage of the available space within a finite time, the upper integration limit can be set to a finite value. This results in (derivation in the Supplementary Information, section S3
)
\begin{equation}\label{equ:AnaModel}
    N =  \frac{\kappa I^3}{4v_g}\left( 1-
    \frac{\pi}{12} \right) + \frac{\kappa I^2 t_0}{2} \left( \frac{t_0^2 v_g^2}{3} - 1 \right),
\end{equation}
\adj{where $I$ is a parameter in the integration limit. \adjJ{A lower limit for the number of parasitic crystals can be set, if $I = D$. At this point, the initial crystals will first touch, but do not cover the entire space.} An upper limit can be determined by using the time it takes for the seed crystals to cover the entire available space, \adjJ{$I = 1.15 D$. In this case, parts of the system are considered twice, resulting in an overestimation of the number of crystals.} The limits are sketched in the inset in \autoref{fig:AnModel}. $v_g, \kappa, t_0$ and $\eta$ 
can be measured from a PF simulation of nucleation on a bare substrate (see Supplementary Information, section S3.1).}\\
\adj{In principle, $\eta$ could be calculated from the nucleation rate $\kappa$ and the growth rate of the nucleated crystals $v_g$. However, \adjJ{nucleation models are not precise enough to estimate the} crystal density (see Supplementary Information, section S3.2). Therefore, $\eta$ is extracted from the simulations.}

The results of the analytical model are plotted on top of the simulation results in \autoref{fig:AnModel}. The two curves overlap. Note that with this set of parameters, the impact of the nucleation induction time is comparatively small.

\subsection{Experimental verification for different nucleation densities}
\begin{figure}[h!]
\centering
\includegraphics[width = 0.75\textwidth]{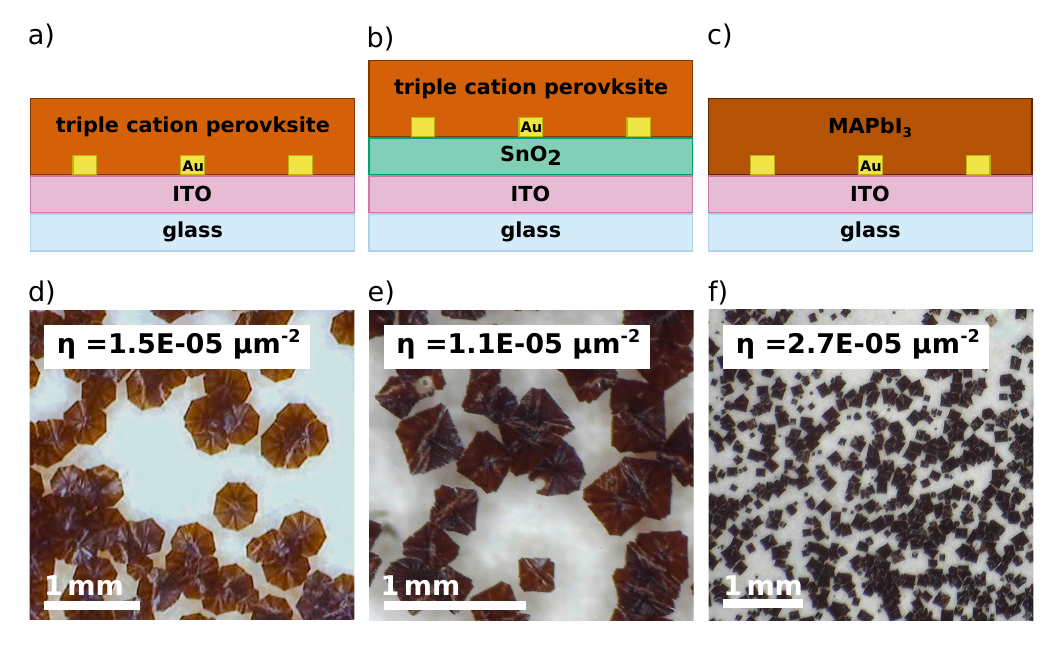}
\caption{Different experimental stacks a), b), c), and directly below the experimental results during recrystallization on unpatterned substrates d), e), f). The different material systems result in different nucleation densities $\eta$.}
\label{fig:DifferentGroups}
\end{figure}
\adj{Both model and simulation are capable of describing the problem at hand with remarkable generality; the only \adj{system specific} parameter 
to the material system is the nucleation density $\eta$ observed on a substrate without seeds. 
To validate the theoretical results, we therefore choose different model systems which feature different values of $\eta$.}\\
Within the experimental setup described in \autoref{secMethylamine}, there are several ways to modify $\eta$. \adjJ{A possible} approach is to vary the atmospheric treatment conditions $T$ and $p_{MA}$: By having the system recrystallize in a more or less pronounced state of supersaturation, the nucleation probability, and thus density, can be shifted substantially. The closer to the phase boundary the system lies in the end, the smaller the nucleation rate, and hence the nucleation density.
\\
Another approach is to vary the material system: The specific substrate/perovskite combination determines the interfacial energy in the liquefied state, thus changing the likelyhood of heterogeneous nucleation. In addition, impurities and roughness at a given substrate surface impact the density of potential nucleation sites, in turn influencing the density of stable nuclei.
\\
To demonstrate the generality of the models presented in section \ref{sec:PhaseFieldSimulationsResults} and \ref{sec:AnalyticalModelResults}, we choose to investigate three different combinations of perovskite composition and substrate material, shown in \autoref{fig:DifferentGroups} a), b), and c), which are chosen strategically to cover a large range of nucleation densities $\eta$.
Two different perovskite compositions are studied: \ce{MAPbI3} and the triple cation perovskite $\ce{Cs}_{0.1}(\ce{MA}_{0.17}\ce{FA}_{0.83})_{0.9}\-\ce{Pb}(\ce{I}_{0.83}\-\ce{Br}_{0.17})_{3}$. Both are deposited on ITO glass, and the triple cation perovksite is additionally studied on a \ce{SnO2} substrate.\\
\newline To perform a seeded growth analysis within the framework of the model presented in section \ref{sec:PhaseFieldSimulationsResults} and \ref{sec:AnalyticalModelResults}, the native nucleation densities $\eta$ have to be extracted for each of the systems. Equivalent to the extraction of $\eta$ from the phase-field simulations in section \ref{sec:PhaseFieldSimulationsResults}, we fabricate films on bare substrates without artificial nucleation centers and let them recrystallize under the same conditions later chosen for the seeded growth. 
Through in-situ microscopic observation of the recrystallization process, the location of each new grain is recorded, thus yielding the values for $\eta$ through a sequential image analysis further described in the methods section. 
Images of the films during recrystallization are shown in \autoref{fig:DifferentGroups} d), e), and f). 
\adj{The resulting values for $\eta$ are \SI{1.1E-5}{\micro\meter^{-2}}, \SI{1.5E-5}{\micro\meter^{-2}} and \SI{2.7E-5}{\micro\meter^{-2}}.}
Based on these values, samples with Au seeds arranged in hexagonal patterns with different distances $D$ are fabricated and analyzed. The selected seed distances $D$ for each system are chosen to cover a range of $D\sqrt{\eta}$ matching the theoretical data in \autoref{fig:AnModel}. The recrystallization process is observed in-situ. Sequential image analysis (further described in the Supplementary Information, section S4) yields the total number of nuclei appearing in the observed area. From there, the number of additional grains per artificial seed is extracted, equivalent to the simulated data in section \ref{sec:PhaseFieldSimulationsResults}.\\
\begin{figure}[h!]
\centering
\includegraphics[width = 0.75\textwidth]{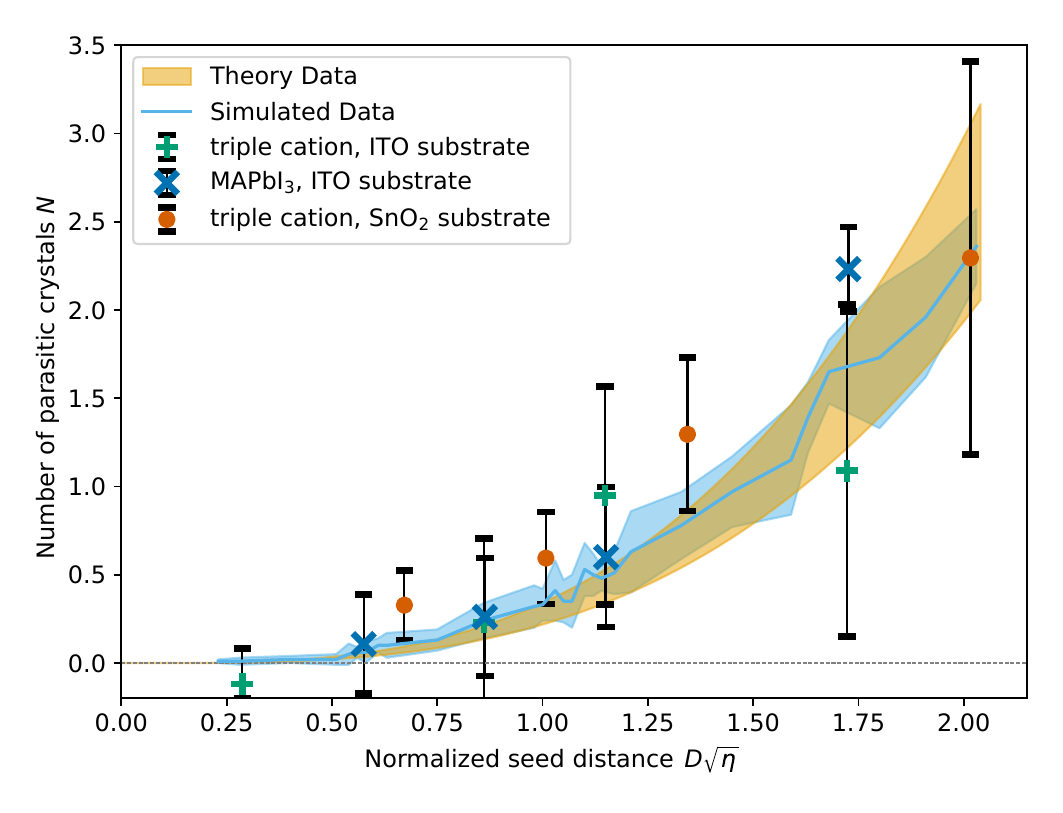}
\caption{Experimental results of the number of excess grains over $D\sqrt{\eta}$ for the three material systems, along with the simulation data and analytical model curves. The experimental results match the theoretical data. The numbers shown here are the mean values derived from multiple growth processes and image analyses. The error bars correspond to the standard deviation of these values. \adjJ{The values result from the analysis of between 6 and 15 image sequences per data point, where each image series tracks one recrystallization process. For more details, reference the methods.}}
\label{fig:FinalResults}
\end{figure}%
The resulting values are shown in \autoref{fig:FinalResults}. The experimental results are in \adjJ{good} agreement with both simulation and the analytical model. The error bars shown represent the standard deviations of the resulting values for all evaluated samples for this pitch.
\\
\newline \adj{Note that, for very low pitches, sometimes fewer grains nucleate than seeds are present on the substrate: In contrast to the simulation, where crystals are placed at these spots initially, nucleation does not occur at all seeds simultaneously. 
It may occur that a grain nucleates and overgrows the neighboring
seeds before nucleation has happened there.
In that case not every seed will lead to a separate grain. This occasionally gives rise to a negative number of \adjJ{parasitic} grains at the low $D\sqrt{\eta}$.\\
Furthermore, the size of the gold seeds has to be considered carefully. If the critical radius for nucleation is smaller than the seeding spot, multiple nucleation events may happen at a single seed. This behavior can be observed for some crystals in \autoref{fig:DirectComparison} d). It will ultimately not impact the growth dynamic itself, and be very hard to recognize in the final film, but a 'grain' started like this will consist of at least two separate crystalline domains.}\\%
\begin{figure}[h!]
\centering
\includegraphics[width = 0.75\textwidth]{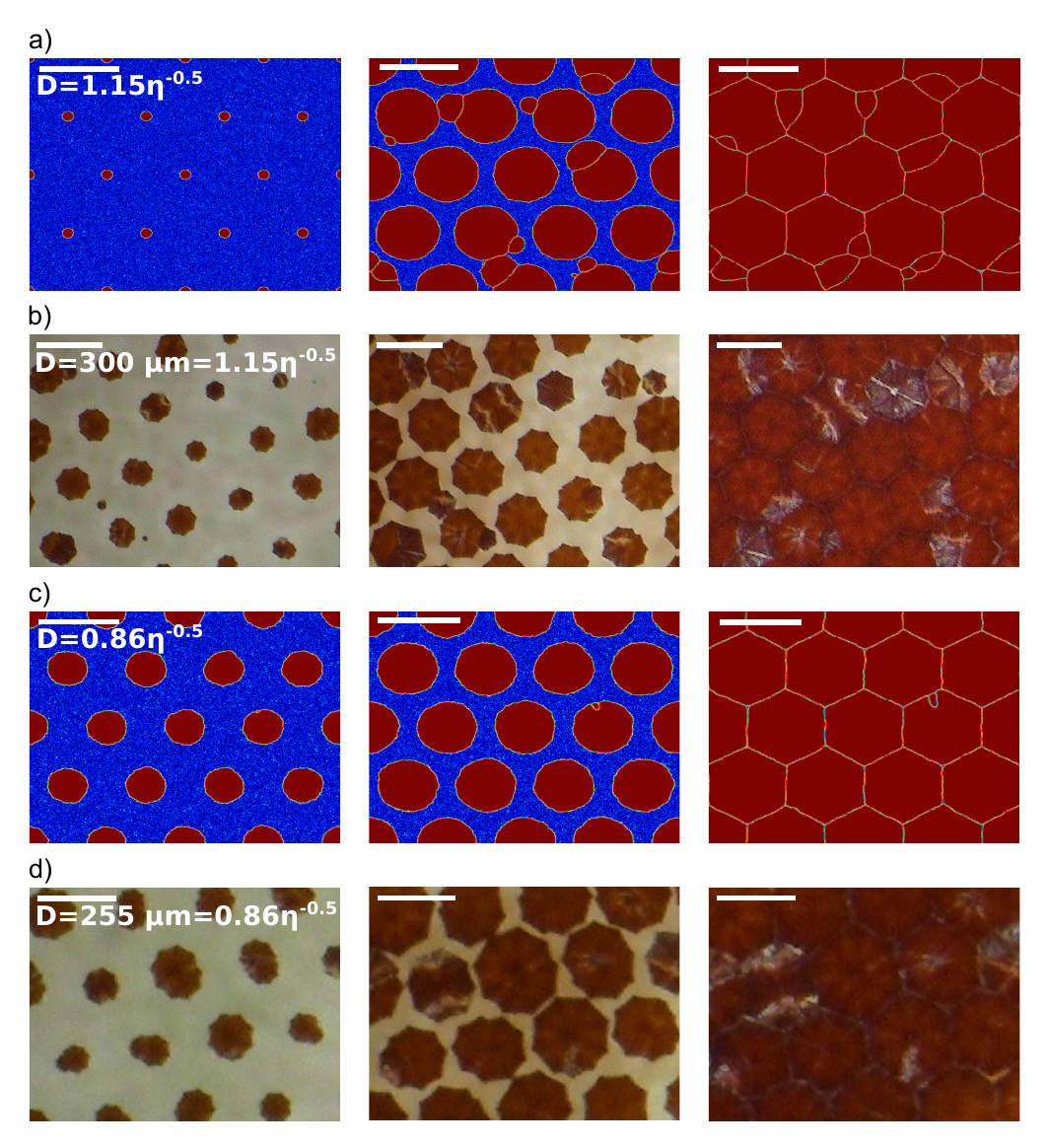}
\caption{
Simulated growth a) and experiment b) for large seed pitches $D$ that lead to the formation of surplus grains. In contrast, c) and d) show simulation and an experiment for smaller $D$, where only the intended grains grow. The temporal progression is shown from left to right. \esc{The bright spots and lines seen in the experimental images on the crystals are caused by light reflections.}}
\label{fig:DirectComparison}%
\end{figure}%
Despite these differences, the model predicts the experiments remarkably well, as is visually shown in \autoref{fig:DirectComparison}.
\adj{Both in the experiments and the simulations, the underlying hexagonal seed pattern is clearly visible. For higher pitches $D$, where $D\sqrt{\eta} \approx 1.2$ (see \autoref{fig:DirectComparison} a) and b), some parasitic nucleation in between the predetermined spots occurs, while that is not the case for the lower pitches shown in \autoref{fig:DirectComparison} c) and d). }

\newpage

\section{Discussion}
The only system specific parameter used in the model is $\eta$ which can be used to adjust the results to a wide range of length scales and defines the possible seed arrangements. 
$\eta$ \adjJ{exists for any material system and} is not exclusive for perovskite systems, hence the prediction should hold for any material system.
Our experiments with different nucleation to growth rate ratios have already shown this flexibility \adjJ{for perovskites. Hence, we postulate that, as long as a continuous film is formed by nearly-isotropic 2D growth of the crystals, our model is able to predict viable distances}. \\
One can, of course, conceive several possible extensions that would open the model to even broader application (see Supplementary Information, section S5).
In its present form, the model already possesses the capability to significantly contribute to the advancement of strategic, highly optimized electronic-- and optoelectronic devices.
In applications and devices where the crystallographic direction and the relative position of grain boundaries are relevant, the model presented here can be used to evaluate possible layouts and to predict limitations of the system in question: For example, if crystals are to be grown across a gap between electrodes (i.e. in transistors  \cite{crystallization_transistors_yu2017,nucleation_mOS2_transistors_han2015seeded}, or photodetectors  \cite{growth_pvsk_photodetectors_wang2015wafer}), we can predict how far apart these electrodes and the nucleation structures can be to ensure single-grain growth as intended.
This \adjJ{idea} can be expanded to nearly all applications where crystals are grown directly onto a functional structure, and where the location, crystallinity, and orientation of the crystal is of importance.
That opens up a large field of research to be explored: At the moment, seeded 2D growth is frequently done from the vapor phase through epitaxy  \cite{epitaxy_wang2016chemically,seeded_epitaxy_xu2021seeded} or vapor deposition  \cite{miseikis2017deterministic}. Solution-based methods are, however, easy to scale up and would be easily integrated in already-established fabrication protocols. Using the understanding and the model we propose here will aid in the systematic planning of experiments and fabrication strategies.
Consequently, this work could aid in expanding on this field in a way that would be highly interesting for industrial production.

\section{Conclusion}\label{sec13}

Millimeter-sized grains can be created with the methylamine treatment process. By adding gold seeds onto the substrate, the spots of nucleation for new grains can be controlled, allowing for full control of size and shape of the crystal. However, parasitic nucleation may occur during this process, potentially spoiling the quality of the final film. To better understand this phenomenon, the system was simulated using a phase-field simulation and an analytical model was developed. These models can predict the occurrence of parasitic nucleation and have been successfully validated against experiments with three different substrate-perovskite material combinations. We find that the value of a dimensionless number, comparing the distance between the seeds and the density of crystals nucleated on a bare substrate defines the occurrence of parasitic nucleation. The prediction of the model is unaffected by the kind of material, the length scale of the system, and the nucleation to growth rate ratio, making it very flexible in its application. Consequently, these models can be utilized for the systematic optimization of nucleation patterns in controlled 2D crystallization as demonstrated here with perovskite films. 

\section*{Supplementary information}
\adj{Simulation data is publicly available via https://doi.org/10.5281/zenodo.17200698. The remaining data that support the findings of this study are included in the supplementary information or available from the corresponding authors upon reasonable request.}

\section*{Acknowledgements}
We acknowledge financial support from the Deutsche Forschungsgemeinschaft (DFG) via the Perovskite SPP2196 programme (Projects No. 506698391 and No. 423660474), the European Commission (H2020 Program, Project 101008701/EMERGE), and CRC1719 Chemprint (Project No. 538767711).

\section*{Conflicts of interest}

The authors declare no conflict of interest.

\section*{Authors Contributions}
E. S. and M. M. conceived the idea. E. S. , M. M., L. S. M., O. R., J. H. designed the project. L. S. M., O. R., J. H. supervised the project. E. S. fabricated the samples and carried out the characterizations. M. M. performed the PF simulations and created the analytical model. E. S. and M. M. wrote the text of the manuscript. L. S. M., O. R., J. H. revised the manuscript. All authors contributed to the discussion and commented on the manuscript.

\newpage
\section{Methods}
\subsection{Sample preparation}
Samples are prepared on \ce{ITO} glass substrates (Luminescence Technology Corp., \SI{15}{\ohm}). After four \SI{30}{\minute}-long steps of ultrasonication (deionized water with detergent, pure deionized water, Acetone, and Isopropanol), the samples are subjected to \SI{7}{\minute} of oxygen plasma treatment.\\
For the samples with a \ce{SnO2} layer as growth interface, a \ce{SnO2} nanoparticle suspension (Alfa Aesar, diluted \SI{1}{\milli\liter}:\SI{4.612}{\milli\liter} in deionized water) is spin-coated at \SI{3000}{rpm} for \SI{30}{\second}, followed by \SI{30}{\minute} annealing at \SI{120}{\celsius}.\\
Samples are then patterned with the \ce{Au} seeds. After spin-coating an adhesive layer (TI Prime, Microchemicals) at \SI{000}{rpm} for \SI{000}{\second} and a \SI{2}{\minute} annealing step at \SI{120}{\celsius}, a roughly \SI{1000}{\nano\meter} thick photoresist (AZ MIR 701 (14 CPS), Microchemicals) layer is spin-coated at \SI{}{} for \SI{}{} and is baked for \SI{90}{\second} at \SI{90}{\celsius}. Illumination of the resist is performed using laser lithography at a dose of \SI{250}{\milli\joule\per\centi\meter\squared}. After illumination and a \SI{90}{\second} bake at \SI{120}{\celsius}, the resist is developed for \SI{60}{\second} (AZ 726 MIF, Microchemicals). After the thermal evaporation of \SI{5}{\nano\meter} of \ce{Cr} and \SI{15}{\nano\meter} of \ce{Au}, and a lift-off in Acetone, the samples are ready for perovskite deposition.\\
A \SI{1.3}{\mol} precursor of the triple cation perovskite is prepared in a \ce{N2}-filled glovebox by dissolving \SI{73.4}{\milli\gram} \ce{PbBr2} (Sigma-Aldrich), \SI{507.1}{\milli\gram} \ce{PbI2} (TCI), \SI{22.4}{\milli\gram} \ce{MABr} (Greatcell Solar), and \SI{172}{\milli\gram} \ce{FAI} (Greatcell Solar) in a $4:1$ mixture of DMF:DMSO and adding \SI{53}{\micro\liter} of a \SI{389.7}{\milli\gram\per\milli\liter} \ce{CsI} (Sigma-Aldrich) solution in \ce{DMSO} per \si{\milli\liter} of the solvent mixture. The finished precursor is filtered in a \SI{0.45}{\micro\meter} PTFE filter just before perovskite deposition.
The perovskite layer itself is spin-coated in \ce{N2} atmosphere. The precursor is dropped dynamically onto the spinning substrate at \SI{1000}{rpm}. The sample is then sped up to \SI{6000}{rpm}, where it remains for \SI{25}{\second}. \SI{20}{\second} into this faster step, \SI{250}{\micro\liter} \ce{CB} anti-solvent are dropped onto the substrate. Finally, the perovskite layer forms during a \SI{40}{\min}-long annealing step at \SI{90}{\celsius}.\\

\begin{figure}[h!]
\centering
\includegraphics[width = 0.85\textwidth]{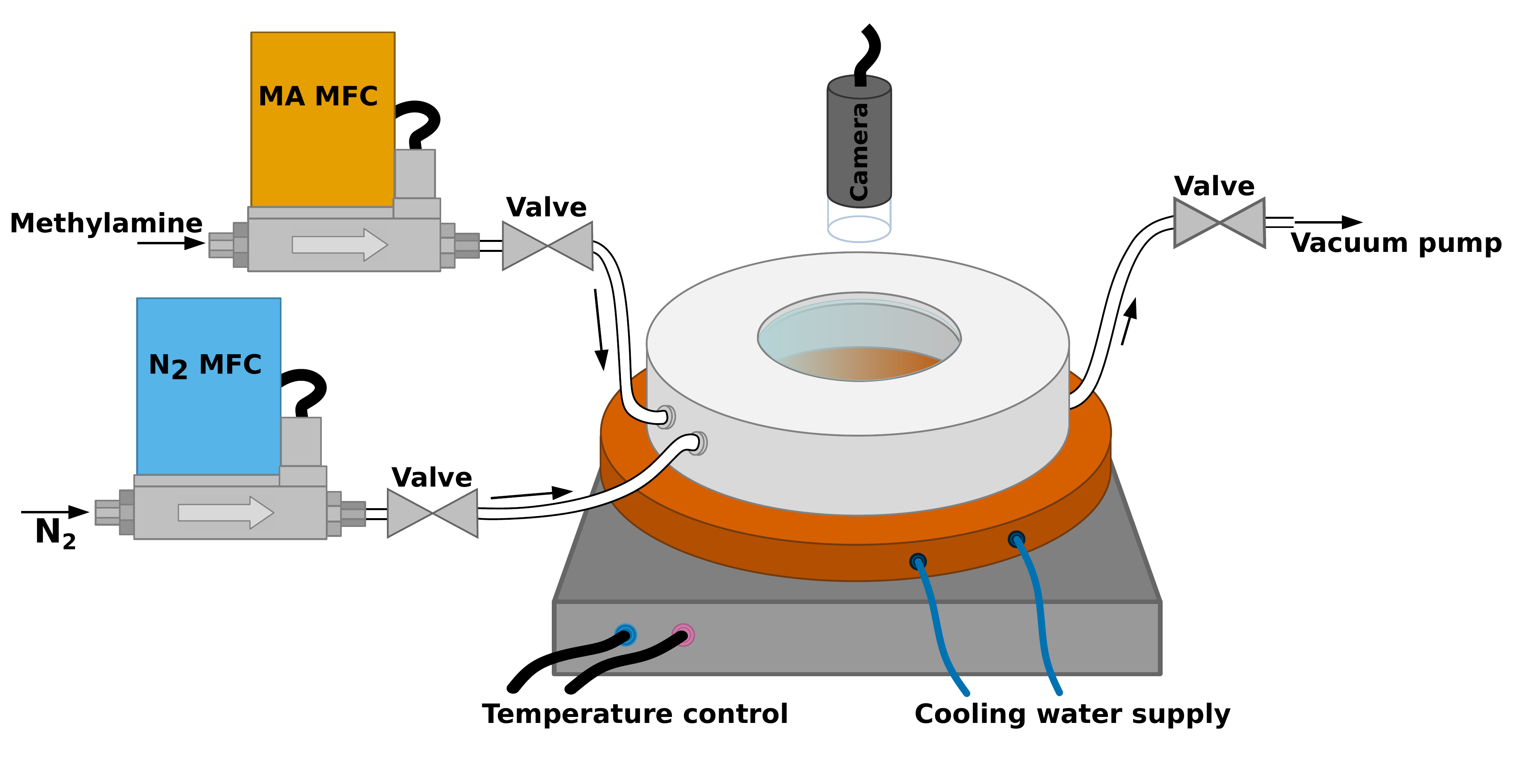}
\caption{Schematic of the methylamine treatment setup. Two mass flow controllers are used to control the amount of \ce{N2} and \ce{MA} gas let into the chamber, which can be heated and cooled from below. A vacuum pump is used to pump to a specific pressure. The process can be observed with a microscope camera through a window of optical glass from the top.
}
\label{fig:SI_setup}
\end{figure}
\subsection{Methylamine treatment setup}
The as-prepared perovskite samples are subjected to methylamine (MA) gas (Linde) in a specially-built automated treatment chamber shown in \autoref{fig:SI_setup}. The chamber can be heated to a specified temperature $T$ with a thermoelement from below. It is equipped with two gas inlets (\ce{N2} and MA) and one outlet, which is connected to a vacuum pump. The inlets are connected to mass flow controllers, allowing for control over the precise amount $N_{MA}$ of gas in the chamber. The total pressure in the chamber is set by pumping to a setpoint below atmosphere pressure $p_{vac}$ with the vacuum pump. Methylamine partial pressures during exposure $p_{MA,1}$ are set by letting a specified amount of gas into the chamber (Volume $V$, temperature $T$):
\begin{align}
    p_{MA,1} = \frac{N_{MA}\cdot R \cdot T}{V}
\end{align}
with the molar gas constant $R$.
The total pressure at this point amounts to $p_{tot,1}=p_{vac}+p_{MA,1}$.
Recrystallization is induced by pumping the chamber down again to $p_{tot,2}=p_{vac}$, thus reducing the methylamine partial pressure $p_{MA,2}$ to:
\begin{align}
    p_{MA,2} = \frac{p_{tot,2}}{p_{tot,1}}\cdot p_{MA,1} = \frac{p_{vac}}{p_{vac}+p_{MA,1}}\cdot p_{MA,1}.
\end{align}
\\
Through an optical glass window over the sample, the entire process is observed in-situ with a microscope camera (Dino-Lite).
\\
\begin{table}[b]
\centering
	\caption{Recrystallization parameters for all three experimental groups. The precise combination of temperature $T$, methylamine partial pressure during recrystallization $p_{MA}$ and substrate-liquid interaction determine the nucleation densities $\eta$ on non-patterned substrates of each system.}\label{table:Methods_settingsForGroups}%
	\begin{tabular}{@{}llll@{}}
		\toprule
			Stack & $T$ & $p_{MA}$ & $\eta$  \\
			\midrule
			Glass/ITO/MAPbI$_3$ & \SI{70}{\celsius} & \SI{195}{\milli\bar} & \SI{2.7E-5}{\micro\meter^{-2}}    \\
			Glass/ITO/3cat & \SI{85}{\celsius} & \SI{230}{\milli\bar} & \SI{1.5E-5}{\micro\meter^{-2}}  \\
			Glass/ITO/SnO2/3cat & \SI{80}{\celsius} & \SI{2.30}{\milli\bar} & \SI{1.1E-5}{\micro\meter^{-2}}  \\
		\bottomrule
	\end{tabular}
\end{table}
The combinations of methylamine partial pressure $p_{MA,2}$ and temperature $T$ are chosen in a way that ensures slow growth of grains and strong preferred nucleation at the \ce{Au} seeds. The settings for all three experiments groups are shown in \autoref{table:Methods_settingsForGroups}.
\subsection{Image analysis and calculation of the number of parasitic crystals}
During the recrystallization process, the sample is observed in-situ from above. From the resulting image material, the total number of nucleated grains $N_{tot}$ in a given observed area $A$ can be extracted. (See the SI for a detailed description of the sequential image analysis employed to that end.) To calculate the number of parasitic grains per artificial seed at a seed pitch $D$, equivalent to the simulated data shown in the paper, we normalize to the primitive hexagonal unit cell area $a=\sqrt{3}D^2/2$. The total number of grains in per unit cell $n_{tot}$ is then calculated via
\begin{equation}
	n_{tot}=\frac{a}{A}\cdot N_{tot}.
\end{equation}
If we had nucleation exclusively at the artificial seeds, we would expect exactly one grain to nucleate in each unit cell. The number of parasitic grains $N$ per unit cell is thus
\begin{equation}
    N=\frac{a}{A}\cdot N_{tot}-1.
\end{equation}
Here, the fraction $\frac{a}{A}$ corresponds to the number of unit cells that fit into the observed area $A$, which is equivalent to the number of seeds in $A$, presuming one seed per unit cell. \\
\begin{table}[t!]
    \centering
	\caption{Number of image sequences analyzed for each combination of pitch and material system.}\label{table:NumberExperimentsPerGroup}%
	\begin{tabular}{@{}llll@{}}
		\toprule
			Stack & Pitch $D$ & normalized Pitch $D\sqrt{\eta}$ & Number of image sequences  \\
			\midrule
			Glass/ITO/MAPbI$_3$ & \SI{50}{\micro\meter} & 0.168 & 8  \\
			Glass/ITO/MAPbI$_3$ & \SI{100}{\micro\meter} & 0.336 & 5 \\
			Glass/ITO/MAPbI$_3$ & \SI{200}{\micro\meter} & 0.672 & 15 \\
			Glass/ITO/MAPbI$_3$ & \SI{300}{\micro\meter} & 1.008 & 12 \\
			Glass/ITO/MAPbI$_3$ & \SI{400}{\micro\meter} & 1.344 & 9 \\
			Glass/ITO/MAPbI$_3$ & \SI{600}{\micro\meter} & 2.016 & 12 \\
			Glass/ITO/MAPbI$_3$ & \SI{1000}{\micro\meter} & 3.360 & 8 \\
            Glass/ITO/3cat & \SI{75}{\micro\meter} & 0.287 & 8 \\
            Glass/ITO/3cat & \SI{150}{\micro\meter} & 0.574 & 8 \\
            Glass/ITO/3cat & \SI{225}{\micro\meter} & 0.861 & 15 \\
            Glass/ITO/3cat & \SI{300}{\micro\meter} & 1.148 & 16 \\
            Glass/ITO/3cat & \SI{450}{\micro\meter} & 1.723 & 8 \\
            Glass/ITO/3cat & \SI{750}{\micro\meter} & 2.871 & 14 \\
            Glass/ITO/SnO2/3cat & \SI{55}{\micro\meter} & 0.288 & 6 \\
            Glass/ITO/SnO2/3cat & \SI{110}{\micro\meter} & 0.575 & 8 \\
            Glass/ITO/SnO2/3cat & \SI{165}{\micro\meter} & 0.863 & 8 \\
            Glass/ITO/SnO2/3cat & \SI{220}{\micro\meter} & 0.150 & 8 \\
            Glass/ITO/SnO2/3cat & \SI{330}{\micro\meter} & 0.173 & 6 \\
		\bottomrule
	\end{tabular}
\end{table}%
In \autoref{table:NumberExperimentsPerGroup}, the number of analyzed images sequences are shown for each combination of pitch and material system shown in \autoref{fig:FinalResults}. Note that the values shown in \autoref{fig:FinalResults} result from the mean and standard deviation of all the values of $N$ extracted for one parameter combination, weighted by the areas $A$ that the analyzed images show for the different image sequences.\\
Note that the number of analyzed image sequences per experimental condition varies. For the smaller pitches in particular, some experiments failed due to contamination or other uncontrolled factors, resulting in fewer analyzable sequences.
\subsection{Determination of phase transition temperatures}
The phase transition points shown in the phase diagram in Figure 1 b) are experimentally determined for triple cation perovskite films. To achieve this, an untreated perovskite film is placed in the chamber and exposed to a defined amount of methylamine gas, corresponding to a methylamine partial pressure $p_{MA}$. This is repeated at increasing temperatures until the film does not liquefy anymore upon methylamine exposure. The corresponding temperature $T$ is then noted as the transition temperature for the partial pressure $p_{MA}$ in question.\\
The precision of the resulting values is limited by the accuracy of the set temperature and that of the set methylamine partial pressure, which is denoted in Figure 1 b) as error bars.

\newpage

\setcounter{section}{0}
\renewcommand{\thefigure}{S\arabic{figure}}
\renewcommand{\thetable}{S\arabic{table}}
\renewcommand{\theequation}{S\arabic{equation}}
\renewcommand{\thesection}{S\arabic{section}}  

\noindent\LARGE{\textbf{Supporting information of 'Controlled nucleation in methylamine-treated perovskite films by artificial seeding and phase-field simulations'}}

\normalsize

\vspace{1cm}

\noindent Emilia R. Schütz$^{a\dagger}$, Martin Majewski$^{b\dagger}$, Olivier J.J. Ronsin$^b$, Jens Harting$^{b,c}$*, Lukas Schmidt-Mende$^a$*\\

\noindent a) Department of Physics, University of Konstanz, Universitätsstraße 10, Konstanz, 78464, Germany\\

\noindent b) Helmholtz Institute Erlangen-Nürnberg for Renewable Energy (HIERN), Forschungszentrum Jülich GmbH\\

\noindent c) Department of Chemical and Biological Engineering and Department of Physics, Friedrich-Alexander-Universität Erlangen-Nürnberg, Fürther Straße 248, Nürnberg\\

\noindent $\dagger$ These authors contributed equally to this work.

\noindent Email Address: j.harting@fz-juelich.de, lukas.schmidt-mende@uni-konstanz.de\\

\section{Experimentally controlling nucleation during the methylamine treatment}

\begin{figure}[h!]
\centering
\includegraphics[width = 0.8\textwidth]{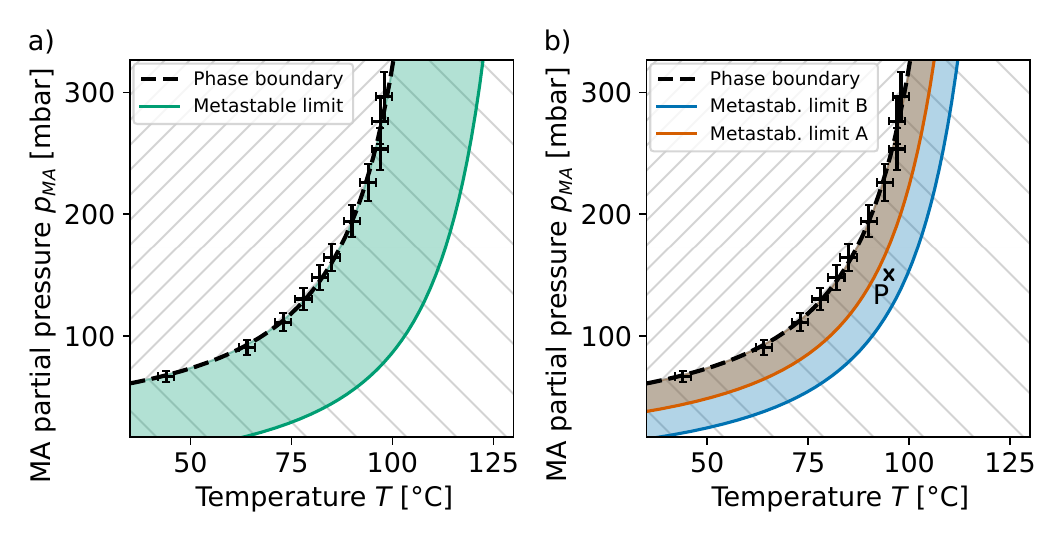}
\caption{a) Phase diagram for the recrystallization process at hand, in case of homogeneous nucleation: beyond the phase boundary, a metastable zone exists (green), where nucleation is unlikely. b) Phase diagram and metastable zones from two different substrate materials that offer nucleation interfaces for heterogeneous nucleation. Depending on the specific interfacial energy with each material, the metastable zones will have different depths, leading to different nucleation probabilities for different materials. If crystallization is initialized at a point $P$, within the metastable zone for one material but not the other, nucleation can be suppressed at the surfaces of material B.
Note that while the phase boundary displayed here consists of experimental data, the metastable limits are qualitative guides to the eye to illustrate the concepts involved and are not based on experimental data.}
\label{fig:PhaseDiagrams}
\end{figure}

In the phase diagram in \autoref{fig:PhaseDiagrams} a), the phase boundary between the solid and liquid states is shown based on experimentally measured data. Left of this curve, the film is liquid in equilibrium, while it is solid in equilibrium on the right side. When moving from the liquid state across the phase boundary, there is a range of temperature-pressure combinations where nucleation is unlikely to occur and the liquid state remains metastable. This  metastable zone is bounded by the metastable limit curve (green line in \autoref{fig:PhaseDiagrams} a)), beyond which nucleation is likely to occur spontaneously.

\begin{figure}[t]
\centering
\includegraphics[width = 0.8\textwidth]{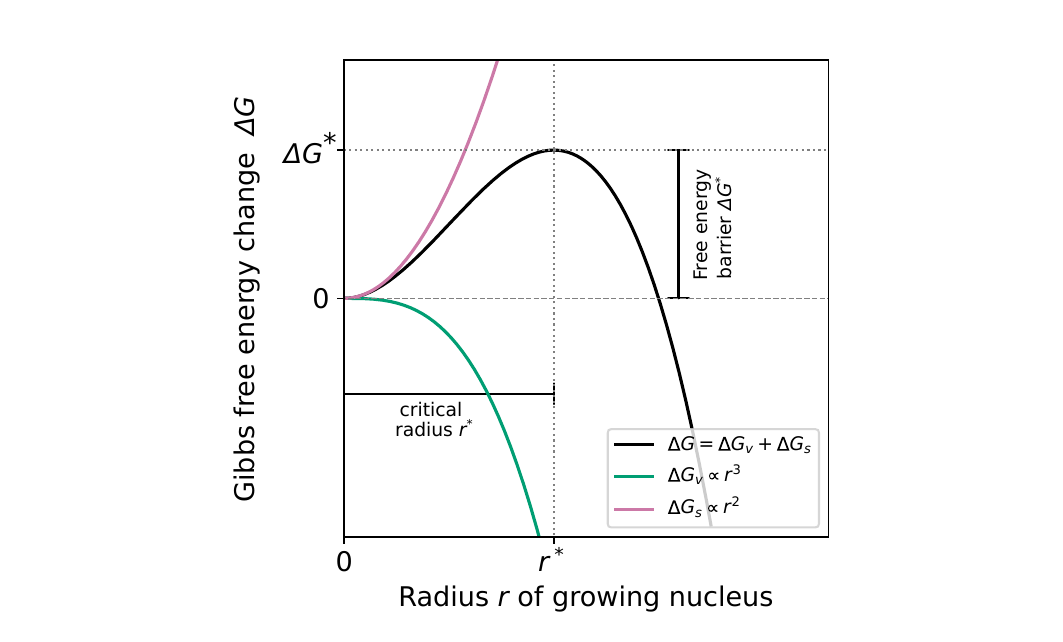}
\caption{Gibbs free energy change $\Delta G$ for the formation of a spherical nucleus in dependence of the nucleus' radius $r$. The formation energy consists of a negative volume component and a positive surface component. The nucleus becomes stable above a critical radius $r^*$, when it has overcome the energy barrier $\Delta G^*$.
}
\label{fig:DeltaGs}
\end{figure}

The presence of the metastable zone and the position of the metastable limit can be understood from classical nucleation theory:\\ The formation of a nucleus is accompanied by a change in Gibbs free energy $\Delta G$.  This free energy change is comprised of a change in the volume free energy $\Delta G_v$, and the surface free energy $\Delta G_s$ \cite{rohtua}:
\begin{equation}
	\Delta G = \Delta G_v+\Delta G_s. \label{equation:deltaG}
\end{equation}
The volume free energy favors nucleation, as the solid phase is more energetically favorable in the region in question. It is the driving force of nucleation and can be expressed as
\begin{equation}
	 \Delta G_v = \frac{4}{3}\pi r^3 \Delta G^*_v\label{equation:deltaGv}
\end{equation}
for a spherical nucleus of radius $r$. $\Delta G^*_v$ is the free energy difference between the liquid and solid state per unit volume. It depends on the degree of supersaturation $S$ and the molecular volume $v$\cite{rohtua}:
\begin{equation}
	\Delta G^*_v = - \frac{k_B\cdot T\cdot ln\left(S\right)}{v}\label{equation:deltaGstarv}
\end{equation}
The surface free energy $\Delta G_s$ is positive, as the formation of an interface between a solid phase and the liquid costs energy. For a spherical nucleus of radius $r$, it can be expressed as
\begin{equation}
	\Delta G_s = 4\pi r^2 \gamma\label{equation:deltaGs},
\end{equation}
with the surface energy per unit area $\gamma$.\\ In \autoref{fig:DeltaGs}, $\Delta G$ is shown against $r$. If a nucleus overcomes the critical nucleus size $r^*$, it becomes stable and can grow into a grain. For that to happen, the energy barrier $\Delta G^*$ has to be overcome. By finding the maximum of \autoref{equation:deltaG}, using \autoref{equation:deltaGv} to \autoref{equation:deltaGs}, this energy barrier can be expressed as \cite{rohtua}
\begin{equation}
	\Delta G^* = \frac{16\pi\cdot \gamma^3}{3\cdot \left(\Delta G_v^*\right)^2}\propto \gamma^3 \propto \left(\Delta G_v^*\right)^{-2}.\label{equation:deltaGstardep}
\end{equation}
\\
$\Delta G^*$ explains the existence of a metastable zone and dictates the position of the metastable limit curve. Within the metastable zone, this energy barrier prevents stable nuclei from forming. Furthermore,  its dependence on $\Delta G_v^*$ directly implies an inverse dependence on the degree of supersaturation $S$. Moving away from the phase boundary, the supersaturation increases. Once $\Delta G^*$ is thus small enough for thermal fluctuations to overcome it, spontaneous nucleation can occur: the metastable limit is reached. \\
 
Thus far, we have only discussed homogeneous nucleation, which theoretically occurs in a medium without any impurities. The only interface, and thus the only interfacial energy of interest, is that between the liquid and solid state. However, in the thin-film system observed here, at least one other interface is possible: that with the substrate. In that case, heterogeneous nucleation occurs: nuclei form at the interface to a third material where the interfacial energy to the growing nucleus is smaller. That decreases the effective interfacial energy because the nucleus does not need to form a full spherical interface with the liquid phase:
It instead  forms a partial interface with (in our case) the substrate, which is energetically favorable. Due to the proportionality to $\gamma^3$ shown in \autoref{equation:deltaGstardep}, this reduction of the effective interfacial energy decreases Gibbs free energy barrier, increasing the likelihood for nucleation. \\
Important to note here is that the \textit{degree} to which the energy barrier decreases depends on the specific interfacial energies between the solid phase, the liquid and the substrate material. Different materials lead to different values of $\Delta G^*$.
In \autoref{fig:PhaseDiagrams} b), the implications for the resulting metastable zones are visualized qualitatively: Effectively, for two different materials, two different metastable limit curves arise. For the material of lower interfacial energy (in our case Au), the metastable zone is smaller, because there is a lower energy threshold to overcome for nucleation, and lower degrees of supersaturation suffice to achieve that. For the material of higher interfacial energy -- in our case ITO -- the metastable zone is larger.
\\
This implies that by controlling $p_{MA}$ and $T$ carefully, we can choose conditions for recrystallization that lie within the metastable zone for nucleation at the ITO interface, but beyond the metastable limit for Au. We, thus, have a high probability for spontaneous nucleation \textit{only} at the Au spots, while nucleation is suppressed everywhere else. 
For larger differences in interfacial energy, the gap between the two metastable limit curves is larger, making is easier to realize an effective nucleation control. Au seems to be particularly well-suited for that purpose \cite{Au_dots_geske2017}, and we attribute that to a favorable interfacial energy between the growing nucleus and the Au surface.
\\
Finally, it is to be noted that while nucleation is unlikely within the metastable zone, it is not impossible. Additionally, impurities and structural inhomogeneities such as larger spikes on the ITO substrate create additional likely nucleation points on the substrates that are not fully accounted for here. Consequently, there is always a degree of spontaneous nucleation in between the seed structures, the limits and implications of which are this paper’s central focus.

\section{Simulation Model}

\subsection{Free Energy}

The simulation code is based on \cite{ronsin_phase-field_2022}. The goal is to investigate crystal nucleation on a patterned substrate. Therefore, only the transition from an amorphous state to a crystalline state is of importance. The evolution of the system is tracked by a crystalline order parameter $\Phi$, which is zero in the amorphous state and becomes one in the crystalline state. To enable the simulation of a polycrystalline system, a so-called orientation parameter $\theta$ is recorded. This labelling field is used to distinguish the different crystals. The Gibbs free energy $G$ is used to record the different energetic contributions:
\begin{equation}
    G = \int_V dV \Delta G= \int_V dV G^{ls} + G^{interf},
\end{equation}
 where $G^{ls}$ describes the liquid-solid phase transition and $G^{interf}$ the interfacial contributions. $G^{ls}$ can be written as

 \begin{equation}
     G^{ls} = \rho\left(\Phi^2(3-2\Phi)L_fus(T/T_m-1) + \Phi^2(\Phi-1)^2 W \right),
 \end{equation}
 with $\rho$ the density of the material, $L_{fus}$ is the enthalpy of fusion, $T$ the temperature, $T_m$ the melting temperature, and $W$ defines the height of the energy barrier upon liquid-solid phase change. The interfacial contribution $G^{interf}$ can be written as
 \begin{equation}
     G^{interf} = \epsilon/2 (\nabla\Phi)^2 + (3-2\Phi)\Phi^2\pi\epsilon_g \delta(\nabla\theta)/2,
 \end{equation}
where $\epsilon$ defines the contribution to the surface energy arising from the gradients in the crystalline order parameter, $\epsilon_g$ for the orientation parameter, and $\delta(\nabla\theta)$ is defined to be one if a difference in orientation is present and zero elsewhere. 

\subsection{Dynamic Equation}

The evolution of the system is given by the stochastic Allen-Cahn equation: 
\begin{equation}
    \frac{\partial \Phi}{\partial t} = - \frac{\nu_0}{RT}M\left(\frac{\partial \Delta G}{\partial \Phi} - \nabla \frac{\partial \Delta G}{\partial (\nabla \Phi)}\right) + d(\Phi,d_\chi,w_\chi,c_\chi) \chi,
\end{equation}
where $\nu_0$ is the molar volume, $R$ the gas constant, $M$ the mobility, $d$ an interpolating function and $\chi$ fluctuations with zero mean and a second moment of
\begin{equation}
    \langle\chi(x,t),\chi(x',t')\rangle = \frac{2\nu_o}{N_a}M\delta(t-t')\delta(x-x'),
\end{equation}
with $N_a$ being the Avogadro number. The interpolating function can be written as
\begin{equation}
    \text{log}(d(d_\chi,w_\chi,c_\chi) = 0.5\, \text{log}(d_\chi)\left( 1 + \text{tanh}(w_\chi(\Phi - c_\chi)) \right),
\end{equation}
whose shape is defined by $w_\chi,d_\chi,c_\chi$. The evolution of the orientation field is given as follows: If the crystalline order parameter rises above a certain threshold $t_\Phi$, then we assign the orientation parameter of the nearest crystal or a new one, if no crystal is present nearby.

\subsection{Simulation Setup}

\begin{figure}[h!]
\centering
  \includegraphics[width = 0.65\textwidth]{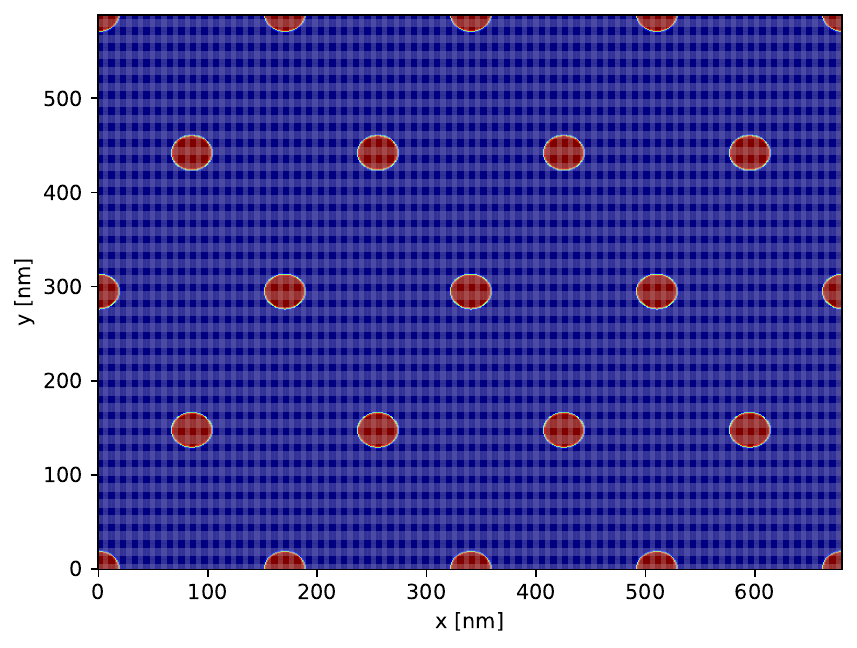}
  \caption{Simulation setup for pitch.}
  \label{fig:Setup}
\end{figure}

The top view of the experiment is simulated. The two-dimensional simulation box has periodic boundary conditions. Sixteen crystals are initially placed in a triangular lattice. The setup is shown in \autoref{fig:Setup}.

\newpage

\subsection{Simulation Parameters}

\begin{table}[h!]
\caption{Parameters of the simulations}
    \centering
        \begin{tabular}{ |p{2.2cm}| p{4.2cm}| p{2.2cm}|p{2.2cm}| }\hline
        \rowcolor{lightgray} Parameters & Full Name & Value & Unit\\ \hline
         dx, dy & Grid spacing & 1 & nm \\  \hline
        nx, ny & Grid size & varied & - \\ \hline
        T & Temperature & 460 & K \\ \hline
        $\rho$ & Density & 1000 & $kg/m^3$ \\ \hline
        m & Molar mass & 0.1 & $kg/mol$ \\ \hline
        $\nu_0$& Molar volume of the Flory Huggins lattice site & $10^{-4}$ & $m^3/mol$ \\ \hline
        $T_m$ & Melting temperature & 600 & $K$ \\ \hline
        $L_{fus}$& Heat of fusion & 75789 & $J/kg$ \\ \hline
        W & Energy barrier upon crystallization & 142105 & $J/kg$ \\ \hline
        $P_0$& Reference pressure & $10^5$ & $Pa$ \\ \hline
        $\epsilon_c$ & Surface tension parameters for order parameter gradient & $1.4\cdot10^{-5}$& $\left(J/m\right)^{0.5}$\\ \hline
        $M_c$& Allen Cahn mobility coefficient for the crystalline phase & 48 & $s^{-1}$ \\ \hline
        $d_{\chi},c_{\chi},w_{\chi}$& Amplitude, center and width of the penalty function for the fluctuations & $10^{-2}$, 0.85, 15 & -\\ \hline
    \end{tabular}
    \label{tab:my_label}
\end{table}

\subsection{Nucleation to growth rate ratio}

\begin{figure}[h!]
\centering
  \includegraphics[width = 0.65\textwidth]{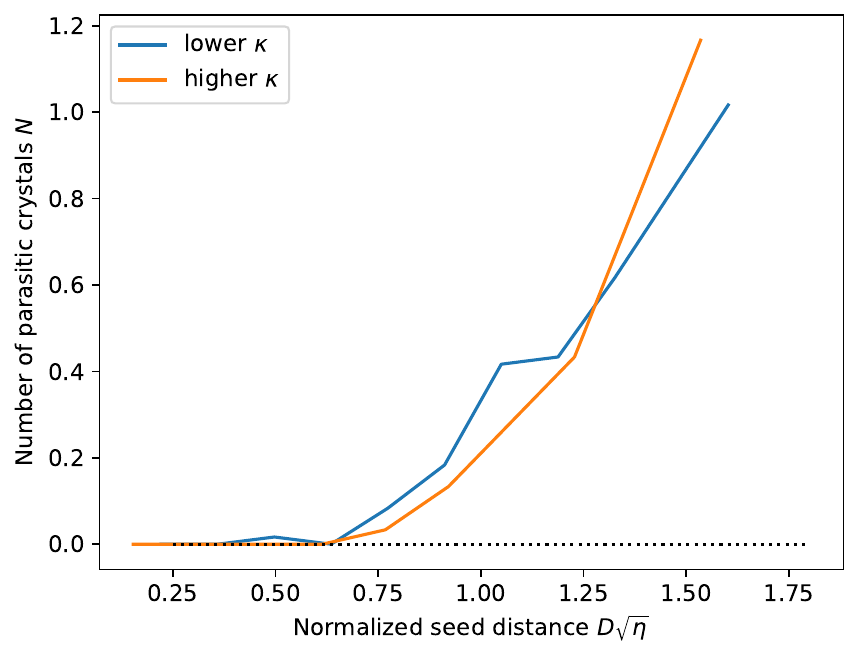}
  \caption{Evaluation of the number of parasitic grains for two different nucleation rates. Since the statistical errors for this process are rather large (see Figure 2, main text), the curves are equivalent, and therefore the ratio between growth and nucleation does not alter the result.}
  \label{fig:nucl}
\end{figure}

The influence of the nucleation-to-growth rate ratio is investigated. To this end, two sets of simulations are performed, varying the seed distance and the nucleation-to-growth rate ratio. As shown in \autoref{fig:nucl}, this ratio leads to changes within the statistical uncertainties. The scaling with the nucleation density $\eta$ on the x-axis is the reason for this behavior.

\subsection{Initial crystal size}

\begin{figure}[h!]
\centering
  \includegraphics[width = 0.65\textwidth]{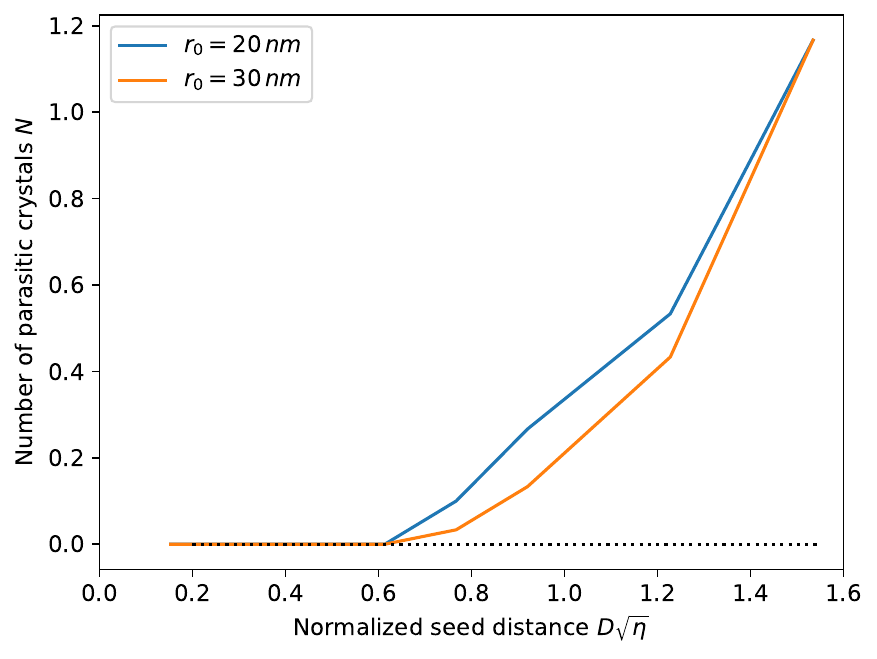}
  \caption{Evaluation of the number of parasitic grains for two different initial crystal sizes. Since the statistical errors for this process are rather large (see Figure 2, main text), the curves are equivalent, and therefore the initial crystal sizes do not alter the result.}
  \label{fig:CrySize}
\end{figure}

The influence of the initial crystal sizes is investigated. To this end, two sets of simulations are conducted, varying the distances between the crystals and the crystal sizes. As shown in \autoref{fig:CrySize}, this size does not matter. The reason is as follows: the distance plotted on the x-axis represents the minimal distance between the surfaces of the crystals, which compensates for the shift in the graph resulting from varied initial crystal sizes.

\newpage
\section{Analytical model}
The experiment utilizes a triangular lattice of gold seeds, where the crystals preferentially nucleate. The dependence of the number of parasitic crystals on the pitch is investigated both numerically and analytically. For the analytical model, the crystal nucleation rate, crystal growth rate, and lag time of nucleation are required as input. These values can be evaluated from the simulations.

The setup is as follows: The area between three seeds is analyzed. The crystal center-to-center distance is called $D$. Therefore, the total area is 
\begin{equation}
    A = D^2/2,
\end{equation}
The time until the crystals impinge on each other is 
\begin{equation}
    t < D/(2v_g).
\end{equation}
It is assumed that crystals can nucleate anywhere within the amorphous phase with a rate of $\kappa$. In the simulation, crystals are placed on a triangular lattice. Since the size of the crystals does not matter (see \autoref{fig:CrySize}), we set the initial size of the crystals to zero in the model, hence:
\begin{equation}
    A_c(t) = (v_g\cdot t)^2\cdot \pi /2
\end{equation}
\adj{Each crystal is part of 6 unit cells in an infinite lattice. Therefore, each of the three circles in the triangle contributes 1/6 of its area.
}

Additionally, there is a nucleation induction time until the first crystal appears. This time is called $t_0$. The total amount of parasitic grains $N$ is therefore
\begin{equation}
    N = \int_{t_0}^{D/2v_g} dt (A - A(t))\kappa.
\end{equation}

\adj{Integrating yield:
\begin{equation}
    N =  \frac{\kappa D^3}{4v_g}\left( 1-
    \frac{\pi}{12} \right) + \frac{\kappa D^2 t_0}{2} \left( \frac{t_0^2 v_g^2}{3} - 1 \right).
\end{equation}
The simplifications used are:
\begin{itemize}
    \item A lower limit is set for $t=D/v_g2$, where the area in between the three impinging circles is still not crystalline.
    \item An upper limit is approximated with $t=1.15 D/v_g2$, where all the material of the system is crystallized due to the growing seeds.
    \item The area covered by nucleated crystals is neglected, and hence, for long seed distances, it overestimates the number of parasitic grains.
\end{itemize} }

\subsection{Analytical model - parameters}
The input parameters to be set are the nucleation rate $\kappa$, the crystal growth rate $v_g$, and the nucleation induction time $t_0$. The crystal growth rate $v_g$ can be measured by simulating one growing crystal. We observe a value of $v_g$ = 142 $nm/s$ for the simulation parameters. For the nucleation rate $\kappa$ and nucleation induction time $t_0$, we use the same simulations, where $\eta$ has been evaluated. We obtain a nucleation induction time of $t_0 = 0.1 s$ and a nucleation rate of $\kappa = 2.1 \cdot 10^{-5} nm^{-2}s^{-1}$. The initial crystal size is set to 0 $nm$, but this value is not relevant for the results, as shown in \autoref{fig:CrySize}.

\subsubsection{Evaluation of the nucleation rate and nucleation induction time}

The nucleation rate and the nucleation induction time are evaluated from a simulation with no initially placed crystals. All further parameters are as stated above. Within the simulation, the nucleation induction time is defined as the time interval from t = 0 to the time at which the first crystal reaches the size of a critical germ. The number of crystals (larger than the critical germ size) and the amount of amorphous area are tracked. Its time derivative is scaled by the (time-dependent) amount of amorphous material to obtain the nucleation rate. This rate is averaged over time to obtain a constant value.

\subsubsection{Crystal growth rate}

The crystal growth rate is measured in a separate simulation, where a single crystal is placed within a box. Fluctuations increase the growth rate compared to the theory (simulations without fluctuations). Therefore, they are turned on. The area of the placed crystal is tracked until it impinges on newly nucleated crystals, and an equivalent radius of the circle is calculated. The growth rate is the time derivative of the radius's evolution.

\subsection{Calculating the nucleation density}

\adj{According to the Johnson-Mehl-Avrami-Kolmogorow (JMAK) theory \cite{van_siclen_random_1996}, the amorphous phase $A(t)$ in a system with homogeneously crystallizing crystals in 2D evolves as
\begin{equation}
    \frac{A(t)}{A_{tot}} = \text{exp}\left(-\frac13 \pi v_g^2\kappa t^3\right),
\end{equation}
where $A_{tot}$ is the total area of the system, $v_g$ the growth rate of the crystals and $\kappa$ the nucleation rate. The nucleation density $\eta$ can be calculated as
\begin{equation}
    \eta = \frac{N}{A_{tot}} = \int_0^\infty {dt A(t)\kappa},
\end{equation}
where $N$ is the total number of crystals in the system. Integrating numerically (with the nucleation and growth rate extracted from the simulation) yields $\eta_{num} \approx 2.7\cdot 10^{13}\,m^{-2}$. The value measured in the simulation ($\eta_{sim} \approx 2.2\cdot 10^{13}\,m^{-2}$) is substantially smaller. This is not unexpected, as JMAK considers only pairwise impingement and therefore overestimates the available area. Hence, we will use the simulated value for higher accuracy.}

\newpage
\section{Sequential image analysis for the detection of the number of nucleating grains}
\begin{figure}[t]
\centering
\includegraphics[width = 0.8\textwidth]{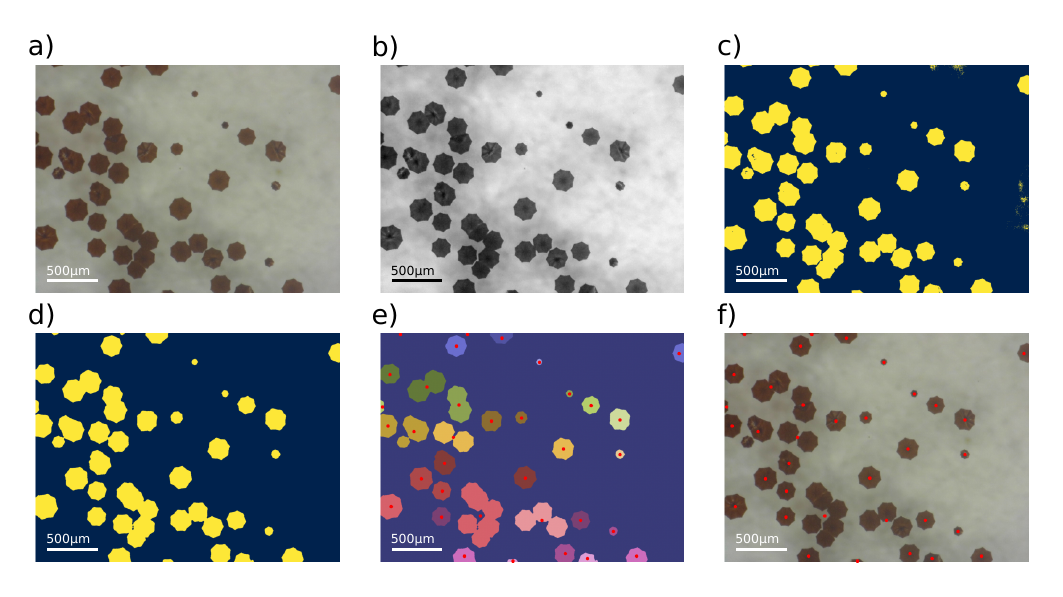}
\caption{Example of the image processing done to determine the centroid points of all grains appearing in one in-situ image. The original image (a) is converted to greyscale (b). The resulting intensities are thresholded (c) to get a binary image. Opening and closing algorithms are used to  get rid of artifacts occurring during thresholding (d). These binary regions are then analyzed (e) to determine the centroid points of each visible grain or grain cluster. The resulting centroids coincide with the grain centers, .i.e, with the nucleation point (f), provided the grain is relatively small and has not yet grown too far or fused with a neighboring grain.
}
\label{fig:ImageAnalysisA}
\end{figure}

\begin{figure}[t]
\centering
\includegraphics[width = 0.8\textwidth]{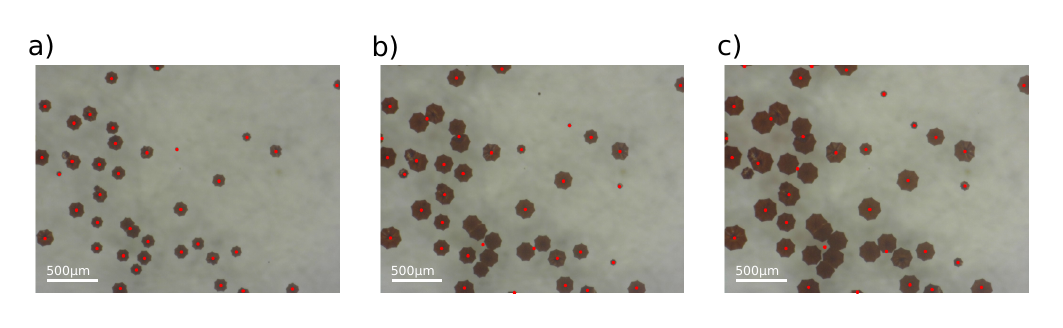}
\caption{Series of images as they are recorded during grain growth with the identified centroid points. While some grains have not touched yet in (a) and generate their own centroid points, they will eventually meet as demonstrated in (b) and (c), resulting in joined centroid points that are not associated with an actual nucleation site.
}
\label{fig:ImageAnalysisB}
\end{figure}

The number of nucleation points per area is established via the image analysis of a sequence of microscope images acquired sequentially during grain growth by a camera placed above the sample. Each image is first analyzed to determine the centroid points of all grains or grain clusters appearing on the sample. Through a comparison of all centroid points appearing on one image to those on the next image, the true nucleation sites are then identified.\\
In \autoref{fig:ImageAnalysisA}, the image processing done to each image is shown. The original image (\autoref{fig:ImageAnalysisA} a)) is first converted to greyscale values (\autoref{fig:ImageAnalysisA} b)). The resulting intensity values are then subjected to thresholding to separate the light background from the dark grains, resulting in a binary image (\autoref{fig:ImageAnalysisA} c)).\\ This binary image will usually still contain some artifacts: ideally, each grain will be a solid area of 'true' color values at this point, and the background will be solidly 'false'. Due to intensity variations on the grains and on the substrate, i.e. due to light reflections, some pixels on the grains will appear too light to be correctly identified as 'true' and vice versa. To eliminate these artifacts, the image is opened, then closed at this point. Opening is a technique used in morphological processing to eliminate small areas below a critical radius, getting rid of noise on the substrate. Closing is the inverse and will fill in missing pixels in the otherwise solid grain regions. A resulting binary image is shown in \autoref{fig:ImageAnalysisA} d). In this image, connected regions of true pixels can be identified (\autoref{fig:ImageAnalysisA} e)). Their size and, crucially, their centroid location can be extracted. The centroid locations extracted this way from the original image are shown overlayed in \autoref{fig:ImageAnalysisA} f).%
\\Analyzing the entire image sequence of grain growth, from the appearance of first nuclei until the substrate is fully covered, the location and number of true nucleation sites can now be determined. As clearly visible in \autoref{fig:ImageAnalysisB} a)-c), the processing of individual images yields the position of the centroid of each grain or grain cluster at that time. In the initial phase of grain growth, this point coincides with the nucleation site, as the grain is still small, grows relatively uniformly in all directions and has not yet fused with any neighboring grains. As the substrate is covered more and more fully, grains meet and fuse into larger regions, whose centroid locations do not correspond to any nucleation points anymore.\\
To identify the true nucleation sites, the centroid positions of each image are compared to those of the next sequentially. If a new centroid appears from one image to the next, the reason has the either be a newly nucleated grain, in which case this point can be noted as a nucleation center, or the meeting of two preexisting grains, which means the new centroid is to be ignored. Conversely, if a centroid disappears from one image to the next, we have to assume the corresponding grain has joined with a neighboring one. 
Upon the joining of two grains with areas $A$ and $B$ and centroids $(x_A,y_A)$ and $(x_B,y_B)$, the new centroid coordinates will be given by:
\begin{align}
    x_{new}=\frac{x_A\cdot A+x_B \cdot B}{A+B}, \qquad y_{new}=\frac{y_A\cdot A+y_B \cdot B}{A+B}\label{equation:centroids}
\end{align}
Therefore, to distinguish between 'real' new nucleation sites and 'false' new joined centroids, we can compare the coordinates of all disappeared old centroids and the areas of the corresponding regions to the coordinates of all new centroids. If two disappeared centroids can be attributed to a new one as described in \autoref{equation:centroids}, the new coordinates can be discarded. Doing this for all images recorded, the true nucleation centers can thus be identified.

\section{Model Extensions}
First, the shape of the patterning seeds may be changed, to reflect and investigate real achievable seeds and patterns. Second, the model could be extended to anisotropic growth mechanisms. As is visible in Fig. 6 (main text), the perovskite films under methylamine atmosphere exhibit preferred growth directions in accordance to the forming crystal lattice. Though this anisotropy is not sufficiently strong in this case to impact the findings from the model, an according extension of the model would be highly interesting for more anisotropic materials.\\
The analytic model and simulations could also be modified to include a third dimension.
Under certain circumstances, there are gaps forming in between the final crystals in the experiment \cite{guenzler_FIRA_2021}. Especially for tight patterns or very slow growth speeds, material may run out before the next grain is met. The origin of this phenomenon would be an interesting starting point for further investigations. To reproduce this phenomenon we expect the simulations have to be more complex by including the third dimension. Additionally, a second material (air) may be necessary to allow for a free surface resulting in pinholes.\\
Furthermore, the simulations and model could be modified to describe a three dimensional system instead of the thin-film based 2D system presented here.\\
Experimentally, patterning systematically in three dimensions would be a challenging endeavor. While one could think of specific scenarios, i.e. inducing nucleation with a light intensity pattern, or mounting nucleation seeds on vertical structures, the experimental effort would be considerable. However, introducing artificial seeds as suspended particles during the fabrication of nanoparticles \cite{He2023_seededGrowth_Nanoparticles} or thin films \cite{Zhao2018_SeedingGrowthPvskCell} is established practice. Such a suspension acts as a statistical distribution of seeds with average distances, and as such could likely be modeled by a similar scheme.

\bibliographystyle{unsrt}
\bibliography{References_tidy}

\end{document}